\documentclass{elsart}
\usepackage{natbib,epsfig,here}
\begin{document}
\runauthor{J\"urgen Schmoll, C.M.Dubbeldam, D.J.Robertson, J.Yao}
\begin{frontmatter}
\title{Optical replication techniques for image slicers}
\author[Durham]{J.Schmoll, D.J.Robertson, C.M.Dubbeldam, J.Yao}
\footnote{Author contact: jurgen.schmoll@durham.ac.uk, Tel. 0044-(0)191-334-4810}
\author[Milano]{F.Bortoletto}
\author[Prague]{L.Pina, R.Hudec}
\author[Marseille]{E.Prieto}
\author[Edinburgh]{C.Norrie, S.Ramsay-Howat}
\author[LFM]{W.Preuss}

\address[Durham]{Center for Advanced Instrumentation (CfAI), Netpark Research Institute,Joseph Swan Road, Netpark, Sedgefield TS21 3FB, United Kingdom}

\address[Milano]{INAF, Vicolo dell'Osservatorio 5, I - 35122 Padova, Italy}
\address[Prague]{Reflex s.r.o., Novodvorska 994, 142 21 Praha 4, Czech Republic}
\address[Marseille]{Laboratoire d'Astrophysique de Marseille, Traverse du Siphon-Les trois Lucs, 13012 Marseille, France}
\address[Edinburgh]{UKATC, Blackford Hill, Edinburgh EH9 3HJ, Scotland}
\address[LFM]{LFM, Badgasteiner Strasse 2, D-28359 Bremen, Germany}

\begin{abstract}
The Smart Focal Planes (SmartFP) activity is a European Joint Research Activity funded to develop novel optical technologies for future large telescope instrumentation \cite{Cunningham2004}. In this paper, we will discuss the image slicer developments being carried out as part of this initiative. Image slicing techniques have many applications in the plans for instrumentation on Extremely Large Telescopes and will be central to the delivery of the science case. A study of a virtual {\it multi-object multi-ifu spectrograph and imager} (MOMSI) for a hypothetical OWL-class telescope reveals the need for focal plane splitting, deployable imagers and very small beam steering elements like deployable IFUs. The image slicer workpackage, lead from Durham University in collaboration with LFM Bremen, TNO Delft, UKATC Edinburgh, CRAL Lyon, LAM Marseille, Padua University and REFLEX Prague, is evaluating technologies for manufacturing micro optics in large numbers to enable multi-object integral field spectroscopy.
\end{abstract}
\end{frontmatter}

\section{MOMSI - a virtual instrument study}
The {\bf MOMSI}  ({\bf m}ulti-{\bf o}bject {\bf m}ulti-ifu {\bf s}pectrograph and {\bf i}mager) project is a virtual instrument, being designed for a 100m class ELT featuring an f/6 focus (\cite{Russell2004_a}, \cite{Russell2004_b}). Due to the physical size of the focal plane (2.92 mm/arcsec, or 350 mm for 2 arcmin FOV), together with the desire to have diffraction-limited sampling, the amount of information available forces a pre-selection of regions of interest. Hence, even for imaging purposes pickoff systems would have to be considered. To get a high multiplex advantage, about 100 deployable IFUs are planned with about 100 $\times$ 100 spatial elements each. The challenges from the micro optics point of view are:
\begin{description}
\item[{\bf Number of sub-systems:}] Each IFU will require a reimaging pickoff system and at least three mirror arrays, being the slicer itself, the pupil and slit mirror. Assuming an array size of 100 elements, there would be 30,000 optical surfaces just for the basic function of a slicer-type IFU. Manufacture and alignment will be critical, as well as the production time scale and costs.
\item[{\bf Size of individual components:}] To reduce the complexity and mass of the whole system, there are no dedicated fore-optics planned for these slicers. Hence the slice width will be in the order of 100 to 300 $\mu$m, being about a factor of 5 less than the typical existing IFU slit widths of e.g. 800 $\mu$m for the GNIRS IFU \cite{All2004}. 
\end{description}

\section{The replication experiments}
Given the prospect of large amounts of similar shaped surfaces, optical replication appears to be a very efficient way of manufacturing tens to hundreds of similar IFUs. Mirror replication techniques are well known for x-ray optics, because for these mirror types classical polishing of the positive forms is very difficult. Several tests were developed to check if this technique is useful for the replication of larger amounts of micro-optics. The replications are performed by two different companies, REFLEX (Prague) and MEDIA LARIO (Lecco). Both groups use roughly the same replication methods, depositing Nickel onto a gold coated mandrel surface. The separation occurs by using the different thermal expansion of the mandrel and replica materials. After that the replica is mounted into a cell of a similar thermal coefficient. The gold layer, that also acts as release agent, sticks to the replica and can be used as an optical surface. Alternatively it can be overcoated with other reflective layers.

\section{Replication trials}
\begin{center}
\begin{figure}[H]
\epsfig{file=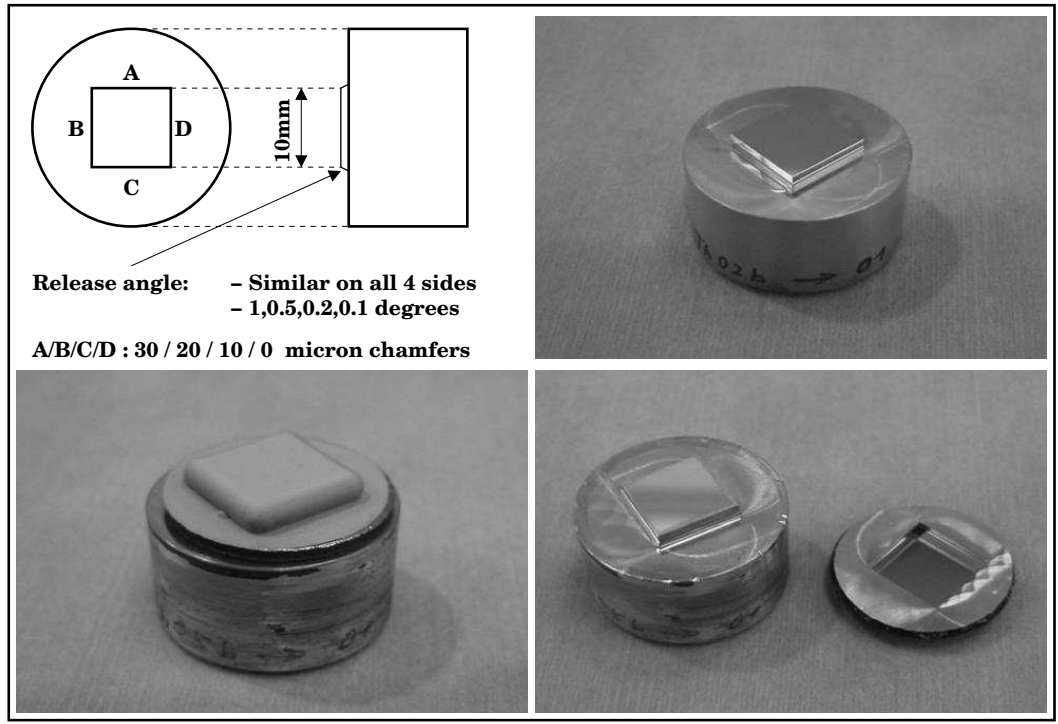, width=140mm}
\caption{The replication stages. Clockwise from upper left: Drawing, gold coated master, after Nickel deposit, separated. The gold layer sticks to the replica surface.}
\label{replicaimages}
\end{figure}
\end{center}

One major issue to address for assessing the suitability of this replication technique was to explore the interface conditions for the master form (mandrel). Usuallly for protruding (male) parts the vertical sides require an angle of at least 1$^{\circ}$ to enable releasing, while the edges require 30 $\mu$m chamfers. While both of these conditions are easy to meet for pupil and slit mirrors, for small slicing mirror components these requirements present problems, especially the chamfer which would lead to unacceptably low fill factors. To determine how these conditions could be optimized for small aperture parts, ``replication challengers'' were produced. These diamond machined aluminium test parts are 10 mm $\times$ 10 mm square, flat surfaces protruding out of a 25mm diameter cylindric blank. Their four edges have chamfers of different size (0, 10, 20 and 30 $\mu$m were specified). Each set consists of four challenger parts, featuring release angles of 1$^{\circ}$, 0.5$^{\circ}$, 0.2$^{\circ}$ and 0.1$^{\circ}$ while having all four chamfer widths each. The split of the lot into four parts was made to assure that at least some parts can be replicated if other fail. Two sets of the four pieces were manufactured by LFM, Bremen, by diamond fly-cutting. Each set was sent to one of the replication companies. Figure \ref{replicaimages} shows a challenger part in the different stages of replication.

\section{Mandrel wear}

The mandrels wear down during replication, making it impossible to reach the desired surface quality after a while. To find out, how long a tool stands, a 100 $\mu$m thick NiP coated challenger part is currently in the process to obtain subsequent replicas from the same mandrel. Additionally REFLEX works with bare aluminium mandrels, using a special type of release agent to detach the parts after replication.

\section{Master material}

The material commonly used is aluminium, which has to be Nickel-Phosphor plated to harden the surface to allow repeated replications. While the aluminium part is easy to produce using diamond machining techniques, the NiP coating tends to introduce surface errors if it is thicker than a few $\mu$m. Alternatively an about 100 $\mu$m thick layer can be applied before machining. Both methods introduce a further complication to make the surface hard enough for subsequent replication. Solutions are 

\begin{description}
\item[{\bf Zerodur mandrels:}] The hardness of Zerodur allows it to be used as a mandrel without the need for NiP coating. Furthermore, the use of Zerodur means that the mandrel-replica sandwich has to be heated up for separation, producing a higher temperature difference than for the standard separation procedure. To test these advantages, two optically polished Zerodur mandrels of the ``challenger'' type have been made to compare their capabilities with the aluminium mandrels. 
\item[{\bf Release agent:}] REFLEX tested a proprietary agent that is deposited onto the mandrel before the gold coating is applied. Due to the improved ease of separating the parts, the mandrel wear is expected to be much lower than for using bare aluminium.
\item[{\bf Double replication:}]
For optical replication the mandrel has to have a negative shape, which can be very difficult to achieve for image slicer components. To solve this problem, the feasibility of a double replication (positive original mandrel, negative replica as intermediate mandrel, and second replica positive again) has been tested by Reflex. Profilometer measurements of the mandrel and both replicas indicate that the deviations introduced by second replication are of the same order as for the single replication.
\end{description}

All three tests are under investigation: While the release agent has been tested at REFLEX already, the Zerodur mandrel replication is currently in progress. The double replication will be checked while replicating pupil mirror prototype parts, which can be machined and tested much easier as a positive shape.

\section{Realistically shaped arrays}

Moving on from the abstract ``challenger'' part to a more realistic facetted surface, a 7-segment array has been designed similar to an all-spherical array used for the GNIRS-IFU pupil mirrors \cite{All2004}. As a positive shape, this array can be compared directly to the replica. This is produced by double replication, testing this process as well as the feasibility of elongated parts with stepped surfaces. First results of REFLEX indicate that double replication of slicing mirror arrays seems to work, albeit quantitative results are not available yet.

\section{First results}

\begin{description}
\item[{\bf Chamfers and angles:}] All chamfer and angle sizes could be reproduced easily, and all parts could be separated.
\item[{\bf Surface roughness:}] The surface roughness does not deteroriate very much, in the case of the release agent used by REFLEX this agent even seems to improve surface roughness, giving rise to smoother replicas.
\item[{\bf Surface shape:}] The shape is subject to a large scale bend, maximum deviations are 2 to 6 $\mu$m. Hence the optical form specifications are not met.
\item[{\bf Process duration:}] 1 mm of Ni layer deposit takes about one week, causing long production cycles. Double replication may be a promising alternative.
\end{description}

While the first two results look promising, shape deformation and process duration can prohibit a use of this technique for the slicer application. Current experiments try to reduce the bending by slightly changing the chemistry of the process. Further experiments are planned to clarify if double replication can be used to reduce the process duration.

\section{Conclusions \& future work}
\vspace{-5mm}
The first results above indicate that the replication of micro-optical surfaces is feasible, even in situations where no large chamfers can be applied. Albeit these results are encouraging, there remain several problems to be solved: Ways to improve the surface roughness and to suppress the surface deformation have to be found. Apart from that, the feasibility to replicate large numbers of pieces within an acceptable time scale has to be proven to exploit the advantages of this technique for the mass production of slicer micro-optics. Currently several ways of double replication are tested to reduce production time.
\vspace{-5mm}

\section{Acknowledgements}

This work is funded by the EU FP6 Opticon Smart Focal Planes consortium.

\end{document}